\documentclass[conference]{IEEEtran}
\IEEEoverridecommandlockouts
\usepackage{amsmath,amssymb,amsfonts}
\usepackage{array}
\usepackage{graphicx}
\usepackage{textcomp}
\usepackage{xcolor}
\usepackage{bm}
\usepackage{bbm}
\usepackage{todonotes} 

\usepackage{multirow}
\usepackage{booktabs}

\usepackage{algorithm}
\usepackage{algpseudocode}

\usepackage[backend=biber,style=ieee,sorting=none, maxbibnames=10]{biblatex}
\usepackage{microtype}

\algtext*{EndWhile}
\algtext*{EndFor}
\algtext*{EndProcedure}
\algtext*{EndIf}
\algtext*{EndSwitch}
\algtext*{EndCase}

\begin{document}
	
	\algnewcommand\algorithmicswitch{\textbf{switch}}
	\algnewcommand\algorithmiccase{\textbf{case}}
	\algnewcommand\algorithmicassert{\texttt{assert}}
	\algnewcommand\Assert[1]{\State \algorithmicassert(#1)}%
	\algdef{SE}[SWITCH]{Switch}{EndSwitch}[1]{\algorithmicswitch\ #1\ \algorithmicdo}{\algorithmicend\ 
	\algorithmicswitch}%
	\algdef{SE}[CASE]{Case}{EndCase}[1]{\algorithmiccase\ #1}{\algorithmicend\ \algorithmiccase}%
	\algtext*{EndSwitch}%
	\algtext*{EndCase}%

\title{Teaching on a Budget in \\Multi-Agent Deep Reinforcement Learning 
}

\author{\IEEEauthorblockN{Erc\"{u}ment \.{I}lhan, Jeremy Gow and Diego Perez-Liebana}
\IEEEauthorblockA{\textit{School of Electronic Engineering and Computer Science} \\
\textit{Queen Mary University of London}\\
London, United Kingdom \\
\{e.ilhan, jeremy.gow, diego.perez\}@qmul.ac.uk}
}

\IEEEpubid{\begin{minipage}{\textwidth}\ \\[12pt]
978-1-7281-1884-0/19/\$31.00 \copyright 2019 IEEE
\end{minipage}}

\maketitle

\begin{abstract}
Deep Reinforcement Learning (RL) algorithms can solve complex sequential decision tasks successfully. 
However, they have a major drawback of having poor sample efficiency which can often be tackled by knowledge reuse. In Multi-Agent Reinforcement Learning (MARL) this drawback becomes worse, but at the same time, a new set of opportunities to leverage knowledge are also presented through agent interactions. One promising approach among these is peer-to-peer action advising through a teacher-student framework. Despite being introduced for single-agent RL originally, recent studies show that it can also be applied to multi-agent scenarios with promising empirical results. However, studies in this line of research are currently very limited. In this paper, we propose heuristics-based action advising techniques in cooperative decentralised MARL, using a nonlinear function approximation based task-level policy. By adopting Random Network Distillation technique, we devise a measurement for agents to assess their knowledge in any given state and be able to initiate the teacher-student dynamics with no prior role assumptions. Experimental results in a gridworld environment show that such an approach may indeed be useful and needs to be further investigated.
\end{abstract}

\begin{IEEEkeywords}
multi-agent, reinforcement learning, deep q-networks, action advising, teacher-student
\end{IEEEkeywords}

\section{Introduction}

Reinforcement Learning (RL) \cite{sutton2018reinforcement} is a prominent framework for solving sequential decision-making tasks.
In recent years, the success of deep learning has led to the emergence of a new set of techniques called deep RL that achieved important breakthroughs, reaching super-human level of play in many complex domains, including Atari games \cite{mnih2015human}, the game of Go \cite{silver2016mastering} and StarCraft II \cite{alphastarblog}.
However, modern RL methods suffer from sample inefficiency and long training times, resulting in limited scalability and applicability in practical situations.
Furthermore, in the cases of having multiple RL agents interacting in a shared environment, which is referred to as Multi-Agent Reinforcement Learning (MARL), these problems are further intensified due to the domain inherent challenges such as non-stationarity and curse of dimensionality introduced by multi-agent dynamics.

One natural solution to accelerate learning in deep RL through overcoming sample inefficiency is \emph{knowledge reuse}.
A considerable amount of methods to leverage past knowledge are proposed to this date \cite{DBLP:journals/jmlr/TaylorS09}.
On the one hand, they are mostly limited to single-agent RL due to the difficulties of MARL.
On the other hand, agent interactions in MARL present unique opportunities of knowledge reuse.

As shown in \cite{da2019survey}, various classes of methods are proposed to take advantage of multiple agents for this purpose, such as imitation learning, inverse RL, and learning from demonstrations. 
\emph{Action advising}, which involves agents exchanging advice in the form of actions between each other to drive their exploration is one of the most flexible approaches among these. The only requirement it has for the agents is to have a common action space and be able to access their local observations through observing or communicating at the time they need an advice. It has a strong track record in single-agent RL and seems to be a promising direction to investigate in MARL.

As Section~\ref{sec:litReview} shows, the research on applications of action advising techniques in MARL is currently in its early stages and the methods are very limited. 
In this study, we aim to investigate the performance of heuristic-based teacher-student framework in cooperative MARL with agents employing nonlinear function approximation representations in their task-level policies. 
The problem setting we address does not hold any assumptions of student-teacher roles for agents, is completely decentralised in training and execution stages, and assumes that the communication between agents is limited; which we believe is realistic.
To the best of our knowledge, this is the first study to utilise classical action advising approach in conjunction with deep MARL, and we hope to shed a light on advantages and shortcomings to determine further research directions.

The paper is structured as follows: Section~\ref{sec:litReview} provides a review of related work. Afterwards, the background information regarding the studied techniques are provided in Section~\ref{sec:back}. Section~\ref{sec:game} describes the game environment used in this project and Section~\ref{sec:method} details the proposed approach. Section~\ref{sec:exp} presents the experimental setup and Section~\ref{sec:res} the results obtained. Finally, Section~\ref{sec:conc} concludes the paper and outlines directions for future work.




\section{Literature Review} \label{sec:litReview}

\subsection{Advising Methods in Single-Agent Reinforcement Learning}

The idea of peer-to-peer knowledge sharing via advising has its roots in single-agent RL.
\cite{clouse1996integrating} is one of the first studies to adopt this method, in a form of student-initiated advising, in which the learning agent is assisted by an expert agent whenever it asks for advice to make a decision.
In \cite{griffith2013policy}, it was shown that it is possible for one of the peers to be a human and provide feedback to accelerate the learning of the agent.
Later on, \cite{DBLP:conf/atal/TorreyT13} proposed the \emph{teacher-student framework}.
According to this method, an expert teacher constantly monitors a student agent's learning process and provides it with action advices at appropriate times, limited with a budget, considering the practical concerns regarding attention and communication.
Addressing the challenge of \emph{when} to advise became more crucial with the introduction of a budget in this line of research.
Following this work, \cite{zimmer2014teacher} treated the optimal way of spending this budget in terms of student's learning performance as an RL problem and attempted to learn how to teach.
In \cite{DBLP:conf/ijcai/AmirKKG16}, a new approach called jointly-initiated is presented. As opposed to the previous methods in which advising occurs with student \cite{clouse1996integrating} or teacher \cite{DBLP:conf/atal/TorreyT13} initiation, they claim that their approach discards the need for student agent to be constantly monitored, making these techniques more feasible for human-agent settings.
\cite{DBLP:journals/corr/ZhanAT16} extended teacher-student framework to take advantage of getting advice from multiple teachers by combining the advices using a voting based selection.
Moreover, the case of getting suboptimal advices when student surpasses teacher's performance is also addressed.
Even though this improvement relaxed the requirement of expert teachers, it still assumes them to follow fixed policies that are good enough to provide advice. 
Finally, \cite{DBLP:journals/corr/FachantidisTV17} made further investigations in learning to teach concept based on \cite{zimmer2014teacher}, and distinguished the qualities of being an expert and being a good teacher, claiming that best performers are not always the best teachers.

All of these previously mentioned studies are based on agents that operate in isolated single-agent environments.
In MARL, multiple agents simultaneously learn in the same environment while affecting each others' policies, rendering it non-stationary.
Due to this, even if we have expert agents, they can no longer be guaranteed to follow a fixed policy. Moreover, the assumption of teacher and student roles within agents are not applicable in MARL, especially with large number of agents. Despite being challenging to adapt on, these properties reflect a more natural and realistic way of peer-to-peer knowledge sharing.

\subsection{Advising Methods in Multi-Agent Reinforcement Learning}

The application of action advising methods in MARL is a challenging and a fairly new subject.
\cite{daSilva:2017:SLA:3091125.3091280} was the first to propose a teacher-student framework that is suitable with multi-agent settings.
They extended the heuristics from \cite{DBLP:conf/atal/TorreyT13} by introducing several metrics based on the number of state visits to measure confidence in a given state, in order to overcome the challenge of having no fixed roles of student and teacher.
They tested their methods in a cooperative team of agents utilizing SARSA(\(\lambda\)) with linear function approximation as task-level policies.
In \cite{DBLP:journals/corr/abs-1805-07830}, teaching in MARL is approached as advising-level meta-learning problem. They proposed a centralised training and decentralised execution procedure using multi-agent actor-critic for teaching-level policies and tabular Q-learning for task-level policies.
Following this work, \cite{kim2019learning} extended the idea of teaching to agents that use deep neural networks in their task-level policies and act in environment with continuous state and action spaces.
Despite having promising results, both of these learning based approaches are currently limited with \(2\) agents only.

\section{Background} \label{sec:back}

\subsection{Multi-Agent RL}

We are interested in a cooperative multi-agent setting where agents get local observations and act in a decentralised fashion. \emph{Decentralised partially observable Markov decision process }(Dec-POMDP)\cite{oliehoek2016concise}, which is a generalisation of \emph{Markov decision process} (MDP) for problems with multiple decentralised agents with local observations, is a suitable formalisation for this purpose. A Dec-POMDP is defined by a tuple $\langle 
\mathcal{I},\mathcal{S},\bm{\mathcal{A}}, 
\mathcal{T},\mathcal{R},\bm{\Omega},\mathcal{O}, \gamma
\rangle$ where $\mathcal{I}$ is the set of $n$ agents, $\mathcal{S}$ is the set of states, 
$\bm{\mathcal{A}} 
= \times_{i \in \mathcal{I}} \mathcal{A}^i$ is the set of joint actions, $\mathcal{T}$ is the state transition 
probabilities, $\mathcal{R}$ is the reward function, $\bm{\Omega} 
= \times_{i \in \mathcal{I}} \Omega^i$ is the finite of set of joint observations, $\mathcal{O}$ is the 
set of conditional observation probabilities, and $\gamma \in [0,1)$ is the discount factor. At each timestep $t$ in a state $s'$, each agent $i$ perceives observation $o^i_t$ determined by $\mathcal{O}(\bm{o} \vert s',\bm{a})$, where $\bm{a} = \langle a^1, \cdots, a^n \rangle$ is the joint action that caused the state transition from $s$ to $s'$ according to $\mathcal{T}(s' \vert \bm{a}, s)$, and receives reward $r^i$ determined by $\mathcal{R}(s,\bm{a}_t)$. In fully-cooperative case, which is the setup we are interested in, every agent gets the same shared reward as $r^i = \cdots = r^n$.
In our problem framework, agents are able to observe each other at any time and infer their local observations. However, they still receive local observations individually from environment. 

\subsection{Deep Q-Networks}

Deep Q-Networks (DQN) \cite{mnih2013playing} is a deep RL algorithm which employs deep neural networks to represent environment states and learn approximations of state-action values similar to Q-learning in an end-to-end fashion. 
It overcomes the limitations present in tabular and linear function approximation methods (like tile coding) which struggle to generalise across states and deal with large state spaces.
At each timestep $t$, an action \(a_t\) is selected randomly (depending on the exploration technique) or greedily according to \(\max_{a} Q(o_t, a ; \omega)\) based on state observation \(o_t\), where \(\omega\) are the parameters of the neural network.
The transition information \(o_t, a_t, r_{t+1}, o_{t+1}\) is then stored in a buffer \(D\), called replay memory.
Using minibatches of samples drawn from \(D\), the network parameters \(\omega\) are trained periodically with gradient descent to minimise the loss \((r_{t+1} + \gamma \max_{a'} Q(o_{t+1}, a') - Q(o_{t}, a_t))^{2}\).

There has been many improvements on DQN over the years \cite{DBLP:conf/aaai/HesselMHSODHPAS18}, and some of them became essential parts of the algorithm as they greatly improve convergence with no significant drawbacks.
In this paper, we utilize double Q-learning, prioritised replay, dueling networks and noisy nets enhancements among these. 

\subsection{Random Network Distillation}

Random Network Distillation (RND) \cite{DBLP:journals/corr/abs-1810-12894} is a technique built to provide an intrinsic curiosity reward for RL agents to enhance their exploration capabilities in complex environments.
It involves usage of two neural networks alongside the actual task-level RL algorithm, namely \emph{target} and \emph{predictor} networks, denoted by differentiable functions \(G\) and \(\hat{G}\) respectively.
These networks are identical in structure, which is defined arbitrarily, and are able to map observations to embeddings as in \(G\colon \mathcal{O} \to \mathbb{R}^{k}\) and \(\hat{G}\colon \mathcal{O} \to \mathbb{R}^{k}\). 
They are initialised with different random weights, and they produce different embeddings for identical inputs in their initial state.
Over the course of training, samples used in task-level RL algorithm are also used to train predictor network \(\hat{G}\) using gradient descent to minimise the mean squared error \(\|G(x) - \hat{G}(x;\theta)\|^{2}\). 
By doing so, the predictor network becomes more accurate at matching target network's outputs for the samples it observes more, which is referred to as distilling a randomly initialised network.
The error between target and predictor are expected to be higher for type of states that are seen less frequently and it acts as a natural indicator of state novelty. 
The authors of \cite{DBLP:journals/corr/abs-1810-12894} state that, due to its simplicity, this method does not have to deal with error types generated by environment stochasticity, model misspecification and learning dynamics, unlike its counterparts that try to predict state transitions.

\subsection{Teaching on a Budget}

The teacher-student framework \cite{DBLP:conf/atal/TorreyT13} is a peer-to-peer knowledge transfer procedure originally developed for two agents, namely the teacher and the student.
While the student is learning to perform a single-agent task, the teacher provides action suggestions to assist its exploration whenever it is appropriate.
These advices are expected to accelerate the student's learning.
Teacher agents also have a form of budget \(b_{give}\) which is the number of advices they can provide to limit the amount they can exchange to reflect realistic scenarios.
The challenge of this framework comes from distributing advices in the most efficient way to accelerate the student's learning progression.


For this purpose, several heuristics are proposed to determine when to advise in teacher-initiated strategy:

\begin{itemize}
	\item \textbf{Early advising:} The intuition is that the earlier states are generally more important in learning, and advice is better taken early on. 
	Following this idea, student is given advice by teacher in the first \(b_{give}\) states.
	
	\item \textbf{Importance advising:} Even though earlier states are usually more important, not all of them may worth to be given advice in. Therefore, a budget is better spent on the states that are determined to be more important. In this method, importance of a state \(s\) is computed as:
	
	\begin{equation}\label{importance}
	I(s) = \max_{a} Q^{teacher}(s,a) - \min_{a} Q^{teacher}(s,a)
	\end{equation}
	
	Advice is provided in a state \(s\) if \(b_{give} > 0\) and \(I(s)\) is greater than the importance threshold \(\tau_{imp}\).

	
\end{itemize}

\section{The Game Environment} \label{sec:game}

\begin{figure}[!t]
    \centering
    \includegraphics[width=.15\textwidth]{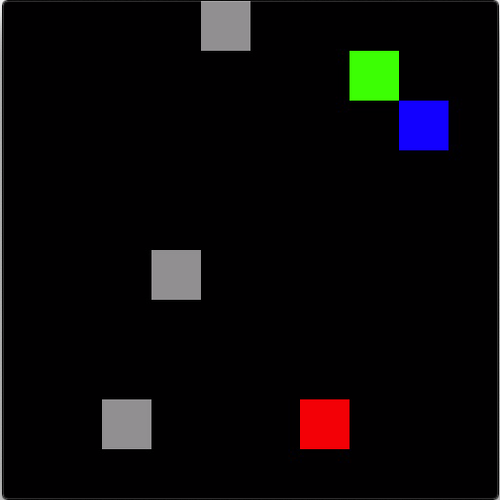}
     \includegraphics[width=.15\textwidth]{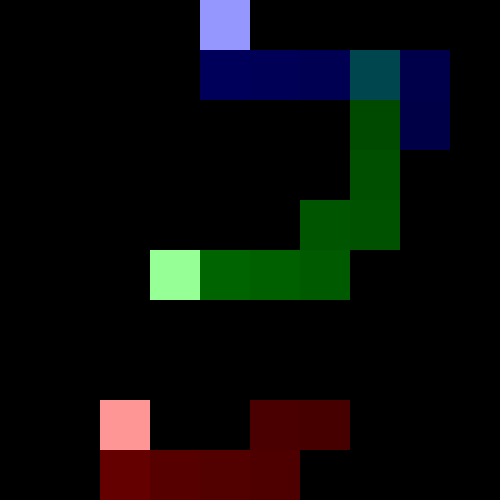}
    \caption{Two frames of the game used for this study. On the left, the initial game state, with three landmarks (grey) and three agents (green, blue and red). On the right, the terminal state with all agents over the landmarks. Agents' trajectories are shown as a shaded colour.}\label{fig:game}
\end{figure}

Our experiment domain is a gridworld game with discrete space and time, in which the agents must spread and cover the landmark positions cooperatively to maximise their shared rewards. The grid is sized $10\times10$ and consists of 3 agents and 3 landmarks. The objective of the agents is to move as quickly as possible to cover all the landmarks of the level. 
While the landmarks are stationary, the agents can navigate around the grid by using a discrete set of actions $\mathcal{A} = \{$ \(Do \: Nothing\), \(Move \: Up\), \(Move \: Down\), \(Move \: Left\), \(Move \: Right\) \(\}\).
Movement actions executed by an agent change its position on the grid by $1$ tile in the corresponding direction.
At each timestep, the agents observe the relative positions of the other agents and landmarks in form of x-axis and y-axis values, plus the current timestep. All values are normalised  to $[-1, 1]$ using the maximum distance and game duration, respectively. Agents also receive a common reward in $[0, 1]$, determined by how many of the landmarks have an agent covering it. The game's reward is calculated as:

\begin{equation}\label{reward}
\frac{1}{|\mathcal{L}|} \sum_{l=1}^{|\mathcal{L}|} \mathbbm{1}_{(\sum_{i=1}^{|\mathcal{I}|} \mathbbm{1}_{\| 
		d_{li} \|} = 0)} \geqslant 1
\end{equation}

where $\mathcal{I}$ is the set of agents, $\mathcal{L}$ is the set of landmarks, $\mathbbm{1}$ is the indicator function, and $d_{li}$ is the Manhattan distance between landmark $l$ and agent $i$. In other words, in order to act optimally and maximise the total reward, the agents must determine the lowest cost (in terms of distance) of the agent-landmark pair set and move to the appropriate landmark following the shortest path.
Since the agents have no access to a forward model, they are expected to learn these strategies through interactions.



 

Despite this environment's representational and mechanical simplicity, it still is capable of presenting complex behavioural challenges for MARL.
Therefore, we decided to conduct experiments in this setting to focus on behavioural learning while keeping the computational expense at minimum. 

We tuned the environment complexity in terms of size and reward sparsity, in order to ensure that the environment is solvable at least sub-optimally by independent DQN agents, and there remains to be some exploration challenge to keep the knowledge of the peers valuable to the agent.

\section{Proposed Method} \label{sec:method}

Our approach aims to accelerate learning of a cooperative team of agents in a multi-agent environment with a teacher-student action advising approach.
The problem setting we take into consideration is fully observable, completely decentralised in every stage, and most importantly, assumes that the environment is too complex for tabular or linear function approximation methods to be successful, unlike the majority of related work.
In this settings, every agent operates with a local observation, however they are able to observe each other and infer any other agent's local observations.
Additionally, they can exchange action advising requests and responses in forms of actions.
We take the budget constraints in inter-agent interactions into consideration as introduced in \cite{DBLP:conf/atal/TorreyT13}, and we want to achieve acceleration while keeping the agent interactions at minimum, since we believe that such behaviour will be the most useful in practical applications. 


The general structure of our proposal can be expressed in two parts: a task-level policy and the teacher-student framework, which we explain in the following sections.

\subsection{Task-Level Policy}

At task-level learning and decision making, agents are designed to employ a nonlinear function approximation according to the target problem structure.
Our agents employ DQN along with its well-established improvements of double Q-learning, prioritised replay, dueling networks and noisy nets.
In our framework, maybe the most important of these is prioritised replay as it helps with sampling transitions produced by advices which are likely to have larger temporal-difference error more frequently. 

The network structure is a multilayer perceptron (MLP), which is the archetypal form of the deep learning models \cite{DBLP:books/daglib/0040158}, with a single hidden layer formed by $256$ units.
All layer connections are dense and noisy.
This addition removes the need to follow an explicit exploration policy such as \(\epsilon\)-greedy while learning, which we think is more suitable for an action advising approach.

As the multi-agent learning strategy, we follow independent learners approach in which each agent treats other agents as part of a non-stationary environment and behaves by taking only its own actions into account.
This is the simplest approach in MARL without requiring the algorithm to have any specialisations for multiple agents. Therefore, we keep DQN as it is without any further modifications in this regard, which is referred to as independent DQNs.
Despite having no theoretical convergence guarantees in multi-agent environments, independent DQNs were able to exhibit promising empirical results in previous studies and are often used as a baseline for further improvements \cite{foerster2016learning}\cite{tampuu2017multiagent}.


\subsection{Teaching on a Budget}

We adopt action advising in the form of the teacher-student framework to perform inter-agent knowledge transfer.
This method requires only a common action set and minimal similarity between agents.
In addition, considering communication costs, exchanging actions instead of episodes or policy parameters is more preferable especially when task-level policy employs a complex model like DQNs.
In our approach, agents can broadcast requests for advices when they need it, and can also respond to requests for advices from other agents.
They have separate budgets of asking \(b_{ask}\) and giving \(b_{give}\) advices, which determine the total number of times they can perform these interactions following either early advising or importance advising heuristics.

Despite its simplicity, adapting the teacher-student framework to MARL with deep RL task-level policies is not straightforward. First of all, in our problem setting, every agent is learning simultaneously and no longer have fixed policies. Moreover, they now have different levels of knowledge about the task, and have no information on each other's expertise since they can not have any access to internal information and they hold on assumptions about roles of being teacher or student. In order to overcome this, we follow a jointly-initiated advising strategy \cite{DBLP:conf/ijcai/AmirKKG16}. Every time an agent requests an advice from another agent, the agent in the position of teacher executes this interaction if it thinks that it can take this role. 
Furthermore, as agents have no fixed roles or knowledge levels that are set previously, they must be capable of determining if they need an advice based on the state they are, or if they are experienced enough to give advice. 
In \cite{daSilva:2017:SLA:3091125.3091280}, they rely on number of visits to measure an agent's certainty in a given state.
However, this is not applicable when state space is large and nonlinear function approximation is used to represent states.
We propose to use RND technique as a metric to measure agents' uncertainties as an alternative for nonlinear function approximation.
Every time an agent uses a sample of state observation to train its internal task-level policy, this sample is also used to \textit{distill} its predictor network \(\hat{G}\).
Consequently, it will be able to measure how uncertain it is in a state with observation \(o\) by measuring the error \(\|\hat{G}(o;\theta) - G(o)\|^{2}\) when determining if it needs advice or is capable of giving it.
This approach can be treated as number of visits in nonlinear function approximation regime.

At each timestep, each agent \(i\) chooses its action according to teacher-student augmented action selection procedure defined in Algorithm \ref{selectaction}.
If it has a remaining asking budget \(b_{ask}\), it measures its uncertainty \(\mu\) using its internal models \(G\) and \(\hat{G}\); if \(\mu\) is higher than a predefined asking threshold \(\tau_{ask}\) the agent broadcasts its advice requests to other agents as described in Algorithm \ref{askforadvice}.
Agents who have any remaining advice giving budget \(b_{give}\) attempt to respond to this request.
If agent receives any advice, it determines which one to follow by using majority voting (ties broken at random), to then execute such action. Otherwise, if no advice is received, it continues exploring by following its own policy.

Responding to an advice request happens as described in Algorithm \ref{advise}. Upon getting a request, the agent first checks if it has any remaining giving budget \(b_{give}\). If so, it then determines its expertise by using its internal \(G\) and \(\hat{G}\), computing \(\|G(o) - \hat{G}(o)\|^{2}\). It then compares this value with threshold \(\tau_{give}\) to decide to proceed or not. Finally, if it is set to follow the Early Advising heuristic, it broadcasts an action advice according to its policy; if the heuristic is set to Importance Advising, it also checks if the state is important enough to give advice for by using Equation \ref{importance}.

The general learning flow of an agent with DQN policy and action advising mechanism is shown in Algorithm \ref{idqnfull}. As can be seen, the task-level policy can easily be isolated from these enhancements except for its action selection policy. Therefore, it can be changed for any other policy as long as it satisfies the assumption of computing state-action values, for importance advising to be applicable.


\begin{algorithm} [!t]
	\caption{Teacher-student augmented action selection}
	\label{selectaction}
	\begin{algorithmic}[1]
	\Require state observation \(o\)
	\Procedure{SelectAction}{}
		\State $a\gets None$
		\If {$b_{ask} > 0$}
			\State $\mu \gets \|G(o) - \hat{G}(o)\|^{2}$ \Comment{Determine uncertainty}
			\If {$\mu > \tau_{ask}$}
				\State $a \gets \Call{AskForAdvice}{o}$	
			\EndIf	
		\EndIf
		\If {$a = None$}
			\State $a\gets \pi(o)$
		\Else
		    \State $b_{ask} \gets b_{ask} - 1$
		\EndIf
		\State \Return $a$
	\EndProcedure
	\end{algorithmic}
\end{algorithm}

\begin{algorithm} [!t]
	\caption{Advice seeking}
	\label{askforadvice}
	\begin{algorithmic}[1]
		\Require state observation \(o\)
		\Procedure{AskForAdvice}{}
			\State \(a\gets None\)
			\State \(A \gets \emptyset\)
			\For{every other agent \(i\)}
				\State \(a_{advice} \gets\) \(i.\)\Call{Advise}{}
				\If {$a_{advice} \not= None$}
					\State add \(a_{advice}\) to \(A\)
				
				\EndIf
			\EndFor
			\If {$A \not= \emptyset$}
				\State $a \gets$ perform majority voting in $A$		
			\EndIf
			\State \Return $a$
		\EndProcedure
	\end{algorithmic}
\end{algorithm}

\begin{algorithm} [!t]
	\caption{Respond to advice request}
	\label{advise}
	\begin{algorithmic}[1]
		\Procedure{Advise}{}
			\State $a\gets None$	
			\If {$b_{give} > 0$}
				\State \(o_{s} \gets\) observe student's state.
				\State $\mu \gets \|G(o_{s}) - \hat{G}(o_{s})\|^{2}$ \Comment{Determine uncertainty}
				\If {$\mu < \tau_{give}$}
					\Switch {teaching method}
						\Case{Early Advising}
							\State $a\gets \pi(o)$
							\State $b_{give} \gets b_{give} - 1$
						\EndCase
						\Case{Importance Advising}
							\State $i_o \gets \max_{a} Q(o_{s},a) - \min_{a} Q(o_{s},a)$
							\If {$i_o > \tau_{importance}$}
								\State $a\gets \pi(o)$
								\State $b_{give} \gets b_{give} - 1$
							\EndIf
						\EndCase
					\EndSwitch
				\EndIf
			\EndIf					
			\State \Return $a$
		\EndProcedure
	\end{algorithmic}
\end{algorithm}

\begin{algorithm} [!t]
	\caption{Training of task-level policy (DQN) with teacher-student framework}
	\label{idqnfull}
	\begin{algorithmic}[1]
		\Procedure{Train}{}
		\State Initialise DQN model
		\State Initialise replay memory
		\State Initialise $G$ and $\hat{G}$
		\For{all training episodes}
			\State $o\gets$ initial observation
			\For{all episode steps}
				\State $a \gets \Call{SelectAction}{o}$
				\State Execute $a$ and get reward $r$ and observation $o'$
				\State Store transition $(o, a, r, o')$ in replay memory
				\State $o \gets o'$	
				\State Sample a minibatch from replay memory			
				\State Train DQN with minibatch
				\State Train $\hat{G}$ with minibatch
			\EndFor
		\EndFor	
		\EndProcedure
	\end{algorithmic}
\end{algorithm}

\section{Experimental Setup} \label{sec:exp}

The objective of the evaluation is to understand if and how the proposed modifications can enhance the learning performance. We conducted experiments through multiple \emph{learning sessions}, each one consisting of a set fixed number of different game episodes which are initialised with random, yet non-overlapping, agent and landmark positions. 


The performance of the agents is assessed as a team through a learning session. They are evaluated every \(100\) episodes in a predefined set of \(50\) evaluation levels.
During evaluation, learning and teaching procedures are disabled, and the levels used are fixed across all learning sessions.
An evaluation score is calculated by normalising the average episode rewards obtained across \(50\) levels with the maximum possible total reward (determined by a set of hand-crafted expert agents), giving a score in $[0, 1]$, where $1$ indicates the optimal performance.

The performance of the proposed methods can be assessed by looking at the evaluation scores across a learning session, according to the following two metrics:

\begin{itemize}
	\item \textbf{Asymptotic performance:} This is measured directly by looking at the evaluation scores values, and represents how good the agents are at solving the game.
	\item \textbf{Learning speed:} This is measured by looking at area under the curve of evaluation scores against the number of training episodes graph. 
\end{itemize}

For agents to be able to benefit from knowledge transfer, there must be some form of knowledge heterogeneity within the team. In MARL, such heterogeneity tends to arise when the agents explore different parts of the state space, use different task-level policies in terms of complexity and representation, and are in different stages of training. Since our environment is fully observable, and we use identical agents, only the latter is applicable in our setting. 

\begin{figure}[!t]
\centering
\includegraphics[width=0.34\textwidth]{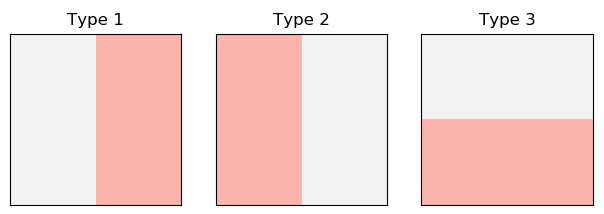}
\caption{Level structure types in terms of possible regions for initial positions of agents (gray) and landmarks (red).}
\label{level-types}
\end{figure}

One objective of this study is to determine how the proposed methods work in different types of knowledge heterogeneity. Therefore, we design the following 2 scenarios:

\begin{itemize}
	\item Scenario I: we train a team of agents in levels from a single distribution of levels; then, we take agents from different stages of pre-training to form a team to be evaluated.
	\item Scenario II: we train 3 sets of agents in 3 different level distributions, in which landmarks and initial agents locations are strictly limited to predefined regions (see Figure \ref{level-types}). Then, we take one agent with moderate performance arbitrarily from each level type to form a team to evaluate. Note that the learning sessions used to pre-train agents are generated with different seeds than the ones where we run the final evaluation.
\end{itemize}

The following is a list of the different types of agents used in this study. These agents are trained in the levels first to acquire some knowledge, and then picked to form teams of 3 agents as explained above. These agents are:

\begin{itemize}
	\item \textbf{A0:} Agent with no prior knowledge.
	\item \textbf{A10:} Agent taken from a team of agents trained for \(10\)K episodes. This agent obtained an evaluation score of \(0.45\) in a level distribution identical to the evaluation sessions.
	\item \textbf{A20:} Agent taken from a team of agents trained for \(20\)K episodes. The evaluation score is \(0.91\) in a level distribution identical to the learning evaluation sessions.
	\item \textbf{A10-1:} Agent taken from a team of agents trained for \(10\)K episodes, with \(0.51\) evaluation score in a restricted level distribution of type 1.
	\item \textbf{A10-2:} Agent taken from a team of agents trained for \(10\)K episodes, with \(0.41\) evaluation score in a restricted level distribution of type 2.
	\item \textbf{A10-3:} Agent taken from a team of agents trained for \(10\)K episodes, with \(0.43\) evaluation score in a restricted level distribution of type 3.
\end{itemize}

The methods we evaluate along with the agent teams used in them are as follows:

\begin{itemize}
	\item \textbf{XP:} The team to be evaluated is formed by agents A0, A10, A20 and no advising is used. This method serves as a baseline for the first scenario.
	\item \textbf{XP-EA:} The team to be evaluated is formed by agents A0, A10, A20, and \texttt{early advising} is enabled.
	\item \textbf{XP-IA:} The team to be evaluated is formed by agents with different knowledge levels A0, A10, A20, and \texttt{importance advising} is enabled.
	\item \textbf{XP-L:} The team to be evaluated is formed by agents A10-1, A10-2, A10-3 and no advising is used. This method serves as a baseline for the second scenario.
	\item \textbf{XP-L-EA:} The team to be evaluated is formed by agents A10-1, A10-2, A10-3, using \texttt{early advising}.
	\item \textbf{XP-L-IA:} The team to be evaluated is formed by agents with different knowledge levels A10-1, A10-2, A10-3, using \texttt{importance advising}.
\end{itemize}

The hyperparameters of the learning algorithms are determined empirically, as shown in Table \ref{table-hyperparameters}. They are kept the same across different scenarios and configurations. 

We run the tests for the methods with teaching enabled, namely XP-EA and XP-IA, with different advice budgets of \(1\)K, \(2\)K, \(5\)K, \(10\)K, \(20\)K and unlimited ($\infty$). For Scenario I (XP, XP-EA, XP-IA), learning is run for \(10\)K episodes, while \(20\)K episodes are run for Scenario II (XP-L, XP-L-EA, XP-L-IA). The budgets are separate for each individual agent and are set the same for asking and giving advice initially. Moreover, due to stability concerns with deep RL methods, especially in multi-agent settings, every experiment is repeated \(10\) times with different random seeds.

\begin{table} [!t]
	\centering
	\caption{Hyperparameters.}
	\label{table-hyperparameters}
	\begin{tabular}{l|r}  
		
		Parameter name & Value \\
  	    \cmidrule(r){1-2}
  	    Minimum replay memory size to start learning & $10000$ \\
		Replay memory capacity & $25000$ \\
		Prioritisation type & proportional \\
		Prioritisation exponent \(\alpha\) & \(0.6\) \\
		Prioritisation importance sampling \(\beta\) & \(0.5 \rightarrow 1\) \\
		Target network update period & $10000$ \\
		Minibatch size & $64$ \\
		Learning rate & $0.001$ \\
		Train period & $2$ \\
		Discount factor \(\gamma\) & \(0.99\) \\
		Adam \(\epsilon\) & \(1.5 \times 10^{-4}\) \\
		Huber loss \(\delta\) & \(1\) \\
		\cmidrule(r){1-2}
		Asking threshold \(\tau_{ask}\) & \(10\) \\
		Giving threshold \(\tau_{give}\) & \(3\) \\
		Importance threshold \(\tau_{imp}\) & \(1\) \\
	\end{tabular}
\end{table}



\section{Results and Discussion} \label{sec:res}

\begin{table*}[!t]
	\centering
	\caption{Asymptotic performance (score) and area under the curve (AUC) values of XP, XP-EA and XP-IA at learning episodes \(2.5\)K, \(5\)K, \(7.5\)K and \(10\)K with standard errors in parentheses. Results of XP-EA and XP-IA that are significantly different than XP algorithm according to Welch's $t$-test with $p < 0.05$ are denoted in bold. Each algorithm is run $10$ times.}
	\label{table-results-sc1}
	\begin{tabular}{lcccccccc}  
	\toprule
	\multirow{2}{*}{Algorithm} & \multicolumn{2}{c}{At 2500th} &  \multicolumn{2}{c}{At 5000th} & \multicolumn{2}{c}{At 7500th} & \multicolumn{2}{c}{At 10000th}\\
	\cmidrule(r){2-3}
	\cmidrule(r){4-5}
	\cmidrule(r){6-7}
	\cmidrule(r){8-9}
	& AUC & Score & AUC & Score & AUC & Score & AUC & Score \\	
	\midrule
	XP 
	& 
	\(13.94 \,(0.04)\) & \(0.58\, (0.005)\) & 
	\(28.72\, (0.12)\) & \(0.62\, (0.009)\) & 
	\(47.15 \,(0.44)\) & \(0.83\, (0.013)\) & 
	\(68.26\, (0.6)\) & \(0.87\, (0.006)\) 
	
	\\
	
	\cmidrule(r){1-9}
	
	XP-EA ($1$K)
	& 
	\(13.9 \,(0.05)\) & \(0.58\, (0.002)\) & 
	\(28.81\, (0.14)\) & \(0.63\, (0.014)\) & 
	\(47.18 \,(0.54)\) & \(0.81\, (0.013)\) & 
	\(68.27 \,(0.73)\) & \(0.86\, (0.01)\) 
	
	\\
	
	XP-EA ($2$K) & 
	\(13.85 \,(0.07)\) & \(0.57 \,(0.005)\) & 
	\(28.58 \,(0.13)\) & \(0.61\, (0.009)\) & 
	\(46.39 \,(0.59)\) & \(0.79\, (0.024)\) & 
	\(67.16 \,(0.87)\) & \(0.86\, (0.009)\) 
	
	\\
	
	XP-EA ($5$K) & 
	\(13.88\, (0.03)\) & \(0.58\, (0.004)\) & 
	\(28.75\, (0.11)\) & \(0.63\, (0.018)\) & 
	\(47.02\, (0.56)\) & \(0.81\, (0.024)\) & 
	\(67.86\, (0.86)\) & \(0.87\, (0.007)\) 
	
	\\	
	
	XP-EA ($10$K) & 
	\(13.85 \,(0.04)\) & \(0.57\, (0.004)\) & 
	\(28.57 \,(0.09)\) & \(0.61\, (0.006)\) & 
	\(46.3 \,(0.47)\) & \(0.8 \,(0.021)\) & 
	\(67.28\, (0.68)\) & \(0.87\, (0.008)\) 
	
	\\
	
	XP-EA ($20$K) & 
	\(13.76 \,(0.05)\) & \(0.57 \, (0.004)\) & 
	\(28.56 \,(0.09)\) & \(0.64 \,(0.016)\) & 
	\(47.38 \,(0.37)\) & \(0.84 \,(0.005)\) & 
	\(68.87 \,(0.45)\) & \(0.88 \,(0.003)\) 
	
	\\
	
	XP-EA ($\infty$) & 
	\(13.76 \,(0.05)\) & \(0.57 \,(0.003)\) & 
	\(28.53 \,(0.06)\) & \(0.64 \,(0.011)\) & 
	\(47.11 \,(0.24)\) & \(0.82 \,(0.006)\) & 
	\(68.0 \,(0.31)\) & \(0.86 \,(0.003)\) 
	
	\\
	
	\cmidrule(r){1-9}

	XP-IA ($1$K) & 
	\(13.94 \,( 0.04) \) & \(0.58 \,( 0.003)\) & 
	\(29.0 \,( 0.15)\) & \(0.66 \,( 0.016)\) & 
	\(47.62 \,( 0.59)\) & \(0.82 \,( 0.012)\) & 
	\(68.92 \,( 0.71)\) & \(0.87 \,( 0.006)\) 
	
	\\
	
	XP-IA ($2$K) & 
	\(13.91 \,( 0.08 )\) & \(0.58 \,( 0.005)\) & 
	\(28.86 \,( 0.21)\) & \(0.64 \,( 0.02)\) & 
	\(47.11 \,( 0.61)\) & \(0.81 \,( 0.015)\) & 
	\(68.16 \,( 0.84)\) & \(0.87 \,( 0.009)\) 
	
	\\
	
	XP-IA ($5$K) & 
	\(13.92 \,( 0.03) \) & \(0.57 \,( 0.004)\) & 
	\(28.84 \,( 0.12)\) & \(\bm{0.65 \,( 0.008)}\) & 
	\(48.01 \,( 0.38)\) & \(0.83 \,( 0.006)\) & 
	\(69.49 \,( 0.41)\) & \(0.88 \,( 0.004)\) 
	
	\\
	
	XP-IA ($10$K) & 
	\(13.97 \,( 0.04) \) & \(0.58 \,( 0.003)\) & 
	\(29.22 \,( 0.22)\) & \(\bm{0.68 \,( 0.024)}\) & 
	\(48.26 \,( 0.72)\) & \(0.8 \,( 0.019)\) & 
	\(69.6 \,( 0.98)\) & \(0.88 \,( 0.006)\) 
	
	\\
	
	XP-IA ($20$K) & 
	\(13.96 \,( 0.04) \) & \(0.58 \,( 0.003)\) & 
	\(29.19 \,( 0.2)\) & \(\bm{0.67 \,( 0.022)}\) & 
	\(47.92 \,( 0.58)\) & \(0.8 \,( 0.012)\) & 
	\(69.01 \,( 0.75)\) & \(0.87 \,( 0.004)\) 
	
	\\
	
	XP-IA ($\infty$) & 
	\(13.96 \,( 0.04) \) & \(0.58 \,( 0.003)\) & 
	\(29.19 \,( 0.2)\) & \(\bm{0.67 \,( 0.022)}\) & 
	\(47.92 \,( 0.58)\) & \(0.8 \,( 0.012)\) & 
	\(69.01 \,( 0.75)\) & \(0.87 \,( 0.004)\) 
	
	\\
	\bottomrule
	\end{tabular}
\end{table*}

\begin{table*}[!t]
	\centering
	\caption{Asymptotic performance (score) and area under the curve (AUC) values of XP-L, XP-L-EA and XP-L-IA at learning episodes \(5\)K, \(10\)K, \(15\)K and \(20\)K with standard errors in parentheses. Results of XP-L-EA and XP-L-IA that are significantly different than XP-L algorithm according to Welch's $t$-test with $p < 0.05$ are denoted in bold. Each algorithm is run $10$ times.}
	\label{table-results-sc2}
	\begin{tabular}{l
			>{\centering\arraybackslash} m{1.2cm}
			>{\centering\arraybackslash} m{1.6cm}
			>{\centering\arraybackslash} m{1.6cm}
			>{\centering\arraybackslash} m{1.6cm}
			>{\centering\arraybackslash} m{1.6cm}
			>{\centering\arraybackslash} m{1.6cm}
			>{\centering\arraybackslash} m{1.6cm}
			c}  
		\toprule
		\multirow{2}{*}{Algorithm} & \multicolumn{2}{c}{At 5000th} &  \multicolumn{2}{c}{At 10000th} & \multicolumn{2}{c}{At 15000th} & \multicolumn{2}{c}{At 20000th}\\
		\cmidrule(r){2-3}
		\cmidrule(r){4-5}
		\cmidrule(r){6-7}
		\cmidrule(r){8-9}
		& AUC & Score & AUC & Score & AUC & Score & AUC & Score \\	
		\midrule
		
		XP-L
		& 
		\(15.38 \,( 0.12)\) & \(0.36 \,( 0.005)\) & 
		\(38.34 \,( 0.43)\) & \(0.52 \,( 0.006)\) & 
		\(66.97 \,( 0.71)\) & \(0.66 \,( 0.02)\) & 
		\(105.46 \,( 1.37)\) & \(0.86 \,( 0.011)\) \\
		
		\cmidrule(r){1-9}
	
		XP-L-EA ($1$K) & 
		\(15.58 \,( 0.1 )\) & \(\bm{0.39 \,( 0.008)}\) & 
		\(\bm{39.51 \,( 0.33)}\) & \(\bm{0.54 \,( 0.005)}\) & 
		\(\bm{69.85 \,( 0.85)}\) & \(0.73 \,( 0.024)\) & 
		\(\bm{110.8 \,( 1.55)}\) & \(0.88 \,( 0.018)\)
		
		\\
		
		XP-L-EA ($2$K) & 
		\(15.4 \,( 0.12 )\) & \(0.37 \,( 0.009)\) & 
		\(38.28 \,( 0.66)\) & \(0.53 \,( 0.012)\) & 
		\(68.55 \,( 1.49)\) & \(0.7 \,( 0.022)\) & 
		\(108.4 \,( 2.28)\) & \(0.86 \,( 0.02)\)
		
		\\
		
		XP-L-EA ($5$K) & 
		\(15.51  \,( 0.14) \) & \(\bm{0.39 \,( 0.007)}\) & 
		\(\bm{39.98  \,( 0.37)}\) & \(\bm{0.54 \,( 0.006)}\) & 
		\(\bm{71.18  \,( 0.68)}\) & \(\bm{0.75  \,( 0.022)}\) & 
		\(\bm{112.69  \,( 1.27)}\) & \(\bm{0.89  \,( 0.005)}\)
		
		\\	
		
		XP-L-EA ($10$K) & 
		\(15.53  \,( 0.1 )\) & \(\bm{0.38  \,( 0.008)}\) & 
		\(\bm{39.53  \,( 0.35)}\) & \(\bm{0.55 \,( 0.005)}\) & 
		\(\bm{71.3  \,( 1.02)}\) & \(\bm{0.74 \,( 0.026)}\) & 
		\(\bm{112.14  \,( 1.77)}\) & \(0.87  \,( 0.014)\)
		
		\\
		
		XP-L-EA ($20$K) & 
		\(15.53  \,( 0.1) \) & \(\bm{0.38  \,( 0.008)}\) & 
		\(\bm{39.53  \,( 0.35)}\) & \(\bm{0.55  \,( 0.005)}\) & 
		\(\bm{71.3  \,( 1.02)}\) & \(\bm{0.74  \,( 0.026)}\) & 
		\(\bm{112.14  \,( 1.77)}\) & \(0.87  \,( 0.014)\)
		
		\\
		
		XP-L-EA ($\infty$) & 
		\(15.53  \,( 0.1 )\) & \(\bm{0.38 \,( 0.008)}\) & 
		\(\bm{39.53  \,( 0.35)}\) & \(\bm{0.55 \,( 0.005)}\) & 
		\(\bm{71.3  \,( 1.02)}\) & \(\bm{0.74 \,( 0.026)}\) & 
		\(\bm{112.14  \,( 1.77)}\) & \(0.87 \,( 0.014)\)
		
		\\
		
		\cmidrule(r){1-9}
		
		XP-L-IA ($1$K) & 
		\(15.39  \,( 0.13) \) & \(\bm{0.38  \,( 0.006)}\) & 
		\(38.54  \,( 0.5)\) & \(0.52  \,( 0.008)\) & 
		\(67.95  \,( 0.97)\) & \(0.68  \,( 0.025)\) & 
		\(107.6  \,( 1.59)\) & \(0.87  \,( 0.01)\)
		
		\\
		
		XP-L-IA ($2$K) & 
		\(15.27  \,( 0.14) \) & \(0.37  \,( 0.008)\) & 
		\(38.4  \,( 0.43)\) & \(0.53  \,( 0.007)\) & 
		\(67.73  \,( 0.69)\) & \(0.67  \,( 0.017)\) & 
		\(106.47  \,( 1.42)\) & \(0.85  \,( 0.016)\) 
		
		\\
		
		XP-L-IA ($5$K) & 
		\(15.37 \,( 0.09) \) & \(0.38 \,( 0.009)\) & 
		\(38.67 \,( 0.32)\) & \(0.52 \,( 0.005)\) & 
		\(68.13 \,( 0.59)\) & \(0.69 \,( 0.025)\) & 
		\(107.32 \,( 1.46)\) & \(0.86 \,( 0.019)\) 
		
		\\
		
		XP-L-IA ($10$K) & 
		\(15.37  \,( 0.1 )\) & \(0.37  \,( 0.009)\) & 
		\(38.58  \,( 0.34)\) & \(0.52  \,( 0.006)\) & 
		\(68.06  \,( 0.65)\) & \(0.68  \,( 0.027)\) & 
		\(107.04  \,( 1.6)\) & \(0.85  \,( 0.02)\)
		
		\\
		
		XP-L-IA ($20$K) & 
		\(15.37  \,( 0.09) \) & \(0.38 \,( 0.009)\) & 
		\(38.67  \,( 0.32)\) & \(0.52  \,( 0.005)\) & 
		\(68.13  \,( 0.59)\) & \(0.69  \,( 0.025)\) & 
		\(107.32  \,( 1.46)\) & \(0.86  \,( 0.019)\) 
		
		\\
		
		XP-L-IA ($\infty$) & 
		\(15.37  \,( 0.09 )\) & \(0.38  \,( 0.009)\) & 
		\(38.67  \,( 0.32)\) & \(0.52  \,( 0.005)\) & 
		\(68.13  \,( 0.59)\) & \(0.69  \,( 0.025)\) & 
		\(107.32  \,( 1.46)\) & \(0.86  \,( 0.019)\)
		
		\\
		\bottomrule
	\end{tabular}
\end{table*}

\begin{figure}[!t]
\centering
\includegraphics[width=3in]{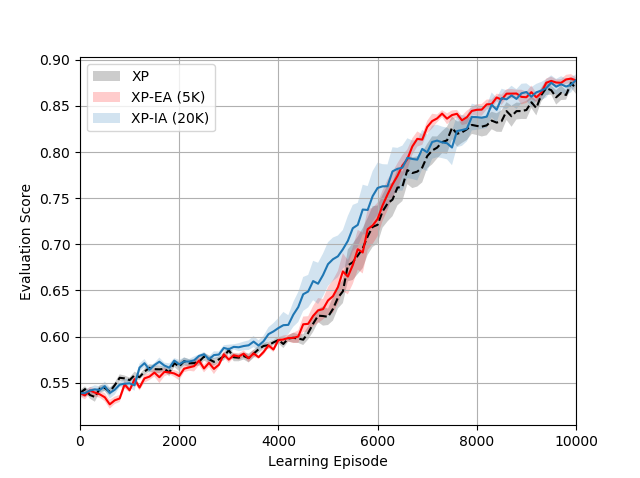}
\caption{Evaluation scores versus number of learning episodes of XP, XP-EA ($5$K) and XP-IA ($20$K). Shaded areas indicate $95\%$ confidence intervals.}
\label{perf1}
\end{figure}

\begin{figure}[!t]
\centering
\includegraphics[width=3in]{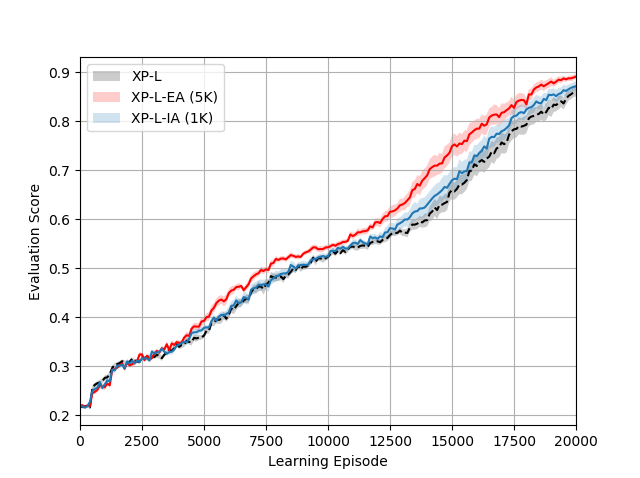}
\caption{Evaluation scores versus number of learning episodes of XP-L, XP-L-EA ($5$K) and XP-L-IA ($1$K). Shaded areas indicate $95\%$ confidence intervals.}
\label{perf2}
\end{figure}

Tables \ref{table-results-sc1} and \ref{table-results-sc2} show the results for the scenarios described in this paper. These results show the asymptotic performance (score) and the area under the curve (AUC) values as indicators of the learning performance of the agents. Results are reported for different moments of the evaluated training: after \(2.5\)K, \(5\)K, \(7.5\)K, \(10\)K, \(15\)K and \(20\)K episodes, and they include the standard error of the measure ($10$ repetitions). Results that are significantly different than the first row (baseline) of each table, according to Welch's $t$-test ($p < 0.05$) are denoted in bold. Additionally, evaluation scores of the methods with highest final AUC values from each of the scenarios are plotted against their respective baseline as shown in Figures \ref{perf1} and \ref{perf2}.

In the first scenario, both of the algorithms provided slight accelerations in learning and achieved very similar final performances with the baseline method at $10$Kth episode. The only significant difference from the baseline was seen at scores at $5$Kth episode, by XP-IA ($5$K). The overall best performing agent in this scenario is XP-IA ($10$K). This can be seen as an indication that the importance metric  was indeed a useful heuristic to distribute advice over more important states in this scenario. The budget seems to have more effect on XP-IA, achieving its best result at 10K, confirming the claim that too much advice may have negative effects on performance \cite{DBLP:conf/atal/TorreyT13}.

In the second scenario, XP-L-IA failed to show any significant advantage over the baseline XP-L except for the score at $5$Kth episode with budget of 1K. This can be a result of the importance metric not being accurate at reflecting actual relevance of states in this kind of agent knowledge setting.
On the other hand, XP-L-EA performed very well with significant improvements in terms of asymptotic performance and learning speed at multiple stages of learning. Moreover, it even managed to achieve a significantly better final performance. This may be caused by the agents starting with similar (and moderate) amount of knowledge, so the early advices are likely to be useful for any of them without having a need for additional importance assessment. The identical results of XP-L with $10$K or a higher budget is caused by not having the need to make use of it beyond some point, once the agent is certain of the decision it is making on its own (as controlled with the advice budgets). This can be considered as an another benefit of using this uncertainty measurement technique if it is tuned well.

\section{Conclusions} \label{sec:conc}

This paper describes the application, for the first time, of action advising via heuristic-based teacher-student framework on Multi-Agent Reinforcement Learning (MARL) agents, employing policies with nonlinear function approximation. The environment used for training agent advising is a grid-based game in which three agents need to coordinate to place themselves in three different landmarks. The work described here shows that using off-policy learning can provide significant improvements in the speed and performance of agents that learn via advice, particularly when the team is composed of agents with heterogeneous knowledge.
Moreover, Random Network Distillation (RND) can be a reliable metric to be utilised as a state visit counter through nonlinear function approximation when state space complexity is high.

Another interesting finding is that the state importance metric may be inefficient in some cases of knowledge distribution amongst agents, for example if they are all are experts but on different state distributions. Additionally, it is worthwhile highlighting that, even if it is possible to determine the experience of the agents for their roles in knowledge exchange relationships, this experience is importantly biased  by the other agents that were present at the time they built their knowledge. Further investigation on how to adapt the importance metric for agent advising is an interesting line of future work.

Off-policy learning through replay memory may be a slowing down factor in action advising, as it takes given advice into consideration in a delayed way and reduces the rate they influence the agent's current policy. Therefore, in addition to the enhancements like prioritised experience replay, it may be useful to implement more specific techniques like multi-step advice and continual monitoring of agents for fixed periods of time after advice exchange. This is similar to previous work in the field \cite{DBLP:conf/ijcai/AmirKKG16}\cite{kim2019learning}.

Finally, another interesting line of future work is to expand the problem to more than $3$ agents, which brings interesting aspects to the discussion such as defining a more accurate peer selection and decision making beyond majority voting.

\printbibliography

\end{document}